\documentclass[aps,PRB,twocolumn,superscriptaddress,preprintnumbers]{revtex4-2}
\usepackage[T1]{fontenc}
\usepackage{times}
\usepackage{graphicx,color}
\usepackage{amsfonts,amsmath,amssymb,amsbsy}
\usepackage[colorlinks=true, linkcolor=blue, citecolor=blue, urlcolor=blue]{hyperref}
\usepackage{float}

\begin{document}

\title{Topological insulators and superconductors based on $p$-wave
magnets,\\
electrical control and detection of a domain wall }
\author{Motohiko Ezawa}
\affiliation{Department of Applied Physics, The University of Tokyo, 7-3-1 Hongo, Tokyo
113-8656, Japan}

\begin{abstract}
Altermagnets are time-reversal broken antiferromagnets, where the $z$
component of the N\'{e}el vector is detectable by anomalous Hall effects. On
the other hand, recently proposed $p$-wave magnets are time-reversal
preserved antiferromagnets, and it is a highly nontrivial problem how to
detect and control a domain wall. We study a one-dimensional hybrid
system made of a $p$-wave magnet and a metal possessing the orbital degree
of freedom. The hybrid system is a topological insulator without the
spin-orbit interaction. There emerge two edge states per one edge, because
the system is mapped to a set of two copies of a topological insulator. Each
copy resembles the long-range Su-Schrieffer-Heeger model but it is
topologically different. Topological interface states emerge at a domain
wall in the $p$-wave magnet, which are charged due to the Jackiw-Rebbi
mechanism. Consequently, a domain wall in the $p$-wave magnet will be
controllable and detectable purely by electrical means. We also study
Majorana fermions induced by proximity coupling of $s$-wave
superconductivity and $p$-wave magnet.
\end{abstract}

\date{\today}
\maketitle




\section{Introduction}

Magnetic domain walls are used for magnetic memories. For example, the
magnetic race-track memory stores information by positions of domain walls%
\cite{Yamaguchi,Parkin}. Readout and control of the position of a domain
wall are essential. In ferromagnetic domain wall, it can be read out by
observing magnetization. In addition, the position of a domain wall is
shifted by applying a current due to the spin-transfer torque\cite%
{Tatara,TataraPR}. On the other hand, it is difficult to read out and
control the position of a domain wall in antiferromagnetic systems\cite%
{Jung,Baltz,Han,Ni,Godin,Kimura,ZhangNeel}.

Altermagnets are the third type of magnets from the viewpoint of symmetry%
\cite{SmejRev,SmejX,SmejX2}. They are expected to be key players of next
generation memories because they have both merits of ferromagnets and
antiferromagnets. Although they are essentially antiferromagnets, the
z-component of the N\'{e}el vector is detectable by measuring anomalous Hall
conductance\cite{Fak,Tsch,Sato,Leiv} because time-reversal symmetry is
broken. The even-parity altermagnets including the $d$-wave, $f$-wave and $i$%
-wave altermagnets are main targets\cite%
{Fak,Tsch,Sato,Leiv,Krem,Lee,Fed,Osumi,Lin,Naka,Gonza,NakaB,Bose}. The band
structure has momentum dependence in altermagnets\cite%
{Ahn,Hayami,Naka,SmejRev,SmejX,SmejX2}, which resembles $d$-wave, $f$-wave
and $i$-wave superconductivity. There are some works\cite%
{Fer,Zu2023,Li2023,Gho,EzawaAlter,LiAlter} on the topological property of
the $d$-wave altermagnet. 

Recently, $p$-wave magnets were proposed\cite{pwave}, where the band
structure has $p$-wave momentum dependence. They have odd parity and
preserve time-reversal symmetry. It is pointed out that CeNiAsO is a $p$%
-wave magnet\cite{pwave}. They are essentially antiferromagnets just as
altermagnets are. Due to time-reversal symmetry, anomalous Hall effects are
absent. It is a highly nontrivial problem to detect and control the magnetic
domain wall in the $p$-wave magnet. The purpose of this work is to overcome
this problem by exploring the topological property based on the $p$-wave
magnet, which is yet to be done. The study of $p$-wave magnets is just a
beginning stage\cite{Maeda}.

In this paper, we study a hybrid system made of the $p$-wave magnet and
one-dimensional metal possessing the orbital degree of freedom. We find the
emergence of two-fold zero-energy topological edge states at each edge. It
is understood by making a unitary transformation, where the system becomes a
set of two copies of one-dimensional topological insulator model. This model
resembles the long-range Su-Schrieffer-Heeger (SSH) model but its
topological properties are different although both models belong to the same
topological class BDI. Next, we show that the Jackiw-Rebbi interface states
emerge at the domain wall in the $p$-wave magnet. Because they are charged,
the domain-wall position will be detectable and controllable
electrically. We also study Majorana edge or interface states induced by
the proximity coupling of $s$-wave superconductivity to a $p$-wave magnet.

\section{$p$-wave topological chain}

\subsection{Model}

The $p$-wave magnet preserves time-reversal symmetry,%
\begin{equation}
\Theta H_{\uparrow }\left( k\right) \Theta ^{-1}=H_{\downarrow }\left(
-k\right) ,
\end{equation}%
where $\Theta =i\sigma _{y}K$\ is the generator of time-reversal symmetry
with $K$\ taking complex conjugate. In the $p$-wave magnet, its effect on
electrons is simply given in the form $\sigma _{z}\sin k$, which satisfies
time-reversal symmetry\cite{pwave}.

We analyze the system where a one-dimensional metal with the orbital degree
of freedom is attached to a $p$-wave magnet. The Hamiltonian is given by 
\begin{align}
H=& \Gamma _{0z}\sum_{x}\frac{t}{2}\left( \hat{c}_{x}^{\dagger }\hat{c}%
_{x+1}+\hat{c}_{x+1}^{\dagger }\hat{c}_{x}\right) -\mu \hat{c}_{x}^{\dagger }%
\hat{c}_{x}  \notag \\
& +\Gamma _{zx}\frac{J}{2i}\sum_{x}\left( \hat{c}_{x}^{\dagger }\hat{c}%
_{x+1}-\hat{c}_{x+1}^{\dagger }\hat{c}_{x}\right) ,  \label{Hamil}
\end{align}%
where $\hat{c}_{x}$ ($\hat{c}_{x}^{\dagger }$)\ is annihilation
(creation) operator of a fermion at the site $x$, $\Gamma _{\mu \nu
}\equiv \sigma _{\mu }\otimes \tau _{\nu }$ is the gamma matrix with $\sigma
_{\mu }$ the Pauli matrix for spins and $\tau _{\nu }$ the Pauli matrix for
orbitals; $t$ is the hopping amplitude and $\mu $ is the chemical potential
for the metallic wire, while $J$ is the magnetization of the $p$-wave
magnet. The meaning of the $p$-wave magnet is clarified in the
momentum-space representation.\ We note that there is no spin-orbit
interaction in the Hamiltonian. The orbital degrees of freedom and their
coupling to magnetization are the same as the Bernevig-Hughes-Zhang model%
\cite{BHZ}. They are studied also in $d$-wave altermagnets\cite%
{Li2023,Gho,EzawaAlter}. The Hamiltonian (\ref{Hamil}) is explicitly
rewritten as%
\begin{eqnarray}
H &=&\sum_{x,\sigma ,\tau }\frac{t}{2}\sigma \left( \hat{c}_{x,\sigma ,\tau
}^{\dagger }\hat{c}_{x+1,\sigma ,\tau }+\hat{c}_{x+1,\sigma ,\tau }^{\dagger
}\hat{c}_{x,\sigma ,\tau }\right)  \notag \\
&&-\mu \hat{c}_{x,\sigma ,\tau }^{\dagger }\hat{c}_{x,\sigma ,\tau }  \notag
\\
&&\!+\frac{J}{2i}\sum_{x,\sigma ,\tau }\tau \left( \hat{c}_{x,\sigma ,\tau
}^{\dagger }\hat{c}_{x+1,-\sigma ,-\tau }-\hat{c}_{x+1,-\sigma ,-\tau
}^{\dagger }\hat{c}_{x,\sigma ,\tau }\right) ,
\end{eqnarray}%
where $\hat{c}_{x,\sigma ,\tau }$ is an annihilation operator of a fermion
with the spin $\sigma $ and the orbital $\tau $,$\ \sigma =\pm 1$\
represents the spin degrees of freedom and $\tau =\pm 1$\ represents the
orbital degrees of freedom.

\begin{figure}[t]
\centerline{\includegraphics[width=0.48\textwidth]{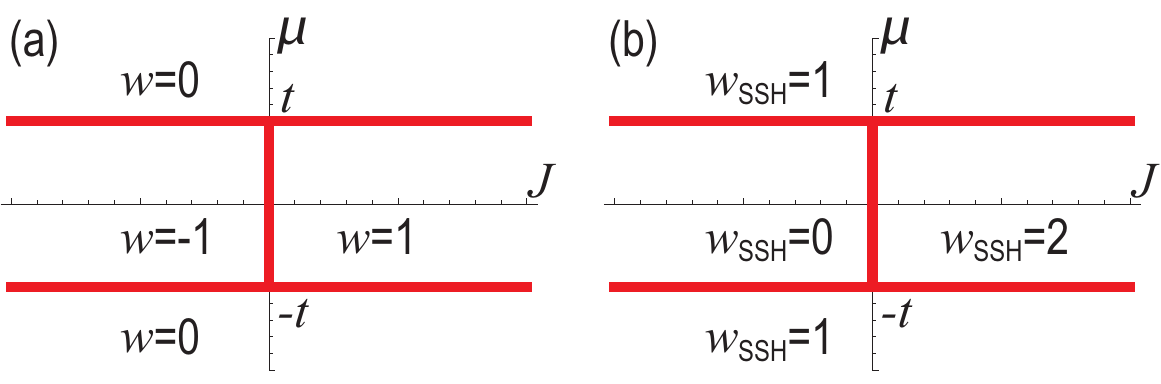}}
\caption{Topological phase diagram in the $J$-$\protect\mu $ plane for (a)
the Hamiltonian $H_{0}$ and (b) the SSH Hamiltonian $H_{\text{SSH}}$. The
gap closes one red lines. The winding number $w$ for the Hamiltonian $H_{0}$
and the winding number $w_{\text{SSH}}$ for the SSH Hamiltonian $H_{\text{SSH%
}}$ are shown in the topological phase diagram.  }
\label{FigPhase}
\end{figure}

The Hamiltonian density reads in the momentum space as%
\begin{equation}
H(k)=\left( t\cos k-\mu \right) \Gamma _{0z}+\left( J\sin k\right) \Gamma
_{zx},  \label{TotalHamil}
\end{equation}%
where $k$\ is the momentum along the $x$-direction, $0\leq k<2\pi $. The $%
\sin k$\ dependence is that of the $p$-wave magnet, which resembles the $p$%
-wave superconductor. Similar models are studied in literature\cite{pwave}%
. The energy spectrum is given by%
\begin{equation}
E(k)=\pm \sqrt{\left( t\cos k-\mu \right) ^{2}+\left( J\sin k\right) ^{2}},
\end{equation}%
where the energy is two-fold degenerated. The gap closing conditions are
given by $\left\vert \mu /t\right\vert =1$\ and $J=0$\ with $\left\vert \mu
/t\right\vert <1$, which are illustrated by\ red lines in the $J$-$\mu $
plane in Fig.\ref{FigPhase}(a).

\begin{figure}[t]
\centerline{\includegraphics[width=0.48\textwidth]{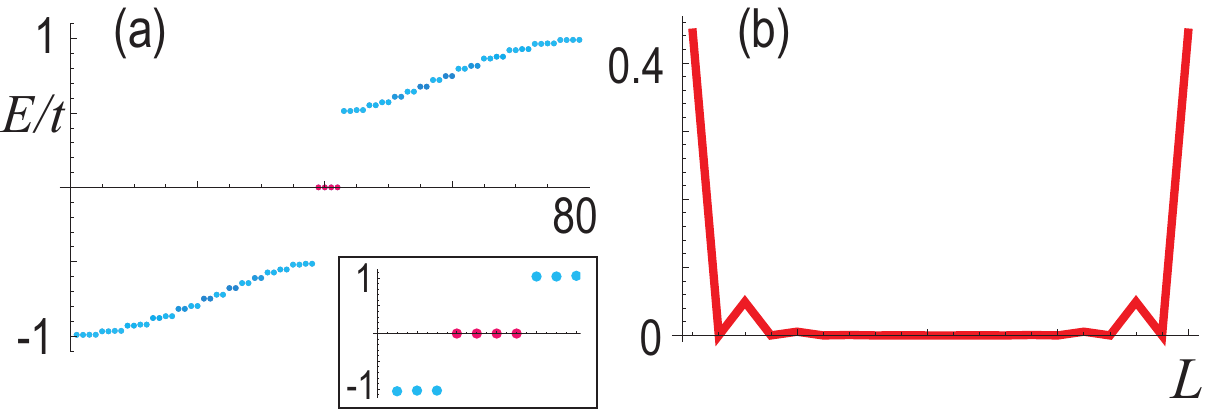}}
\caption{(a) Energy spectrum for a finite chain described by Eq.(\protect\ref%
{Hamil}). The inset shows an enlarged figure in the vicinity of the
zero-energy, where four zero-energy edge states exist. The horizontal axis
is the site index. Color palette for (a) indicates $\left\vert \protect%
\psi _{1}\right\vert ^{2}+\left\vert \protect\psi _{L}\right\vert ^{2}$,
which is given in Fig.\protect\ref{FigDWEne}(d). (b) Spatial profile of the
zero-energy states. We have set $J=0.5t$, $\protect\mu =0$ and $L=20$. }
\label{FigEdge}
\end{figure}

\subsection{Symmetry}

The system has chiral symmetry, $\left\{ H,\Gamma \right\} =0$ with the
chiral operator $\Gamma \equiv \left[ \Gamma _{0z},\Gamma _{zx}\right] /2i$.
It has also time-reversal symmetry, $\Theta H\left( k\right) \Theta
^{-1}=H\left( -k\right) $ with $\Theta =i\sigma _{y}$. In addition, there is
particle-hole symmetry, $\Xi H\left( k\right) \Xi ^{-1}=-H\left( -k\right) $
with $\Xi =\tau _{x}$. Then, the system belongs to the class BDI\cite{Ryu},
whose topology is characterized by the class $Z.$

Chiral symmetry or particle-hole symmetry protects the existence of the
edge states exactly at the zero energy. If we add a symmetry breaking
perturbation term, the edge states may acquire nonzero energy but the
spatial profile of the edge states persists in the first-order perturbation
theory. 

\subsection{Topological edge states}

We study a chain with a finite length. The energy spectrum is shown in Fig.%
\ref{FigEdge}(a). There emerge four zero-energy edge states as in the inset
of Fig.\ref{FigEdge}(a). They are edge states as in Fig.\ref{FigEdge}(b),
where two zero-energy edge states are present at each edge. We show that
they are topological edge states in the following subsection.

\begin{figure}[t]
\centerline{\includegraphics[width=0.48\textwidth]{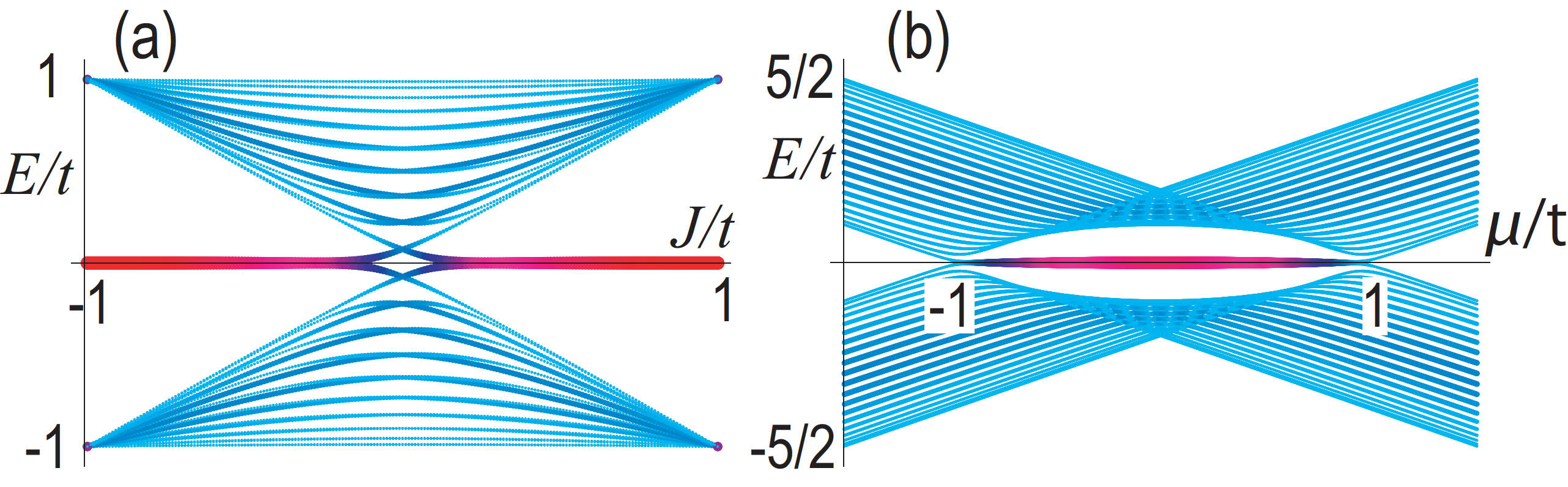}}
\caption{(a) Energy spectrum as a function of $J/t$, where we have set $%
\protect\mu =0$. (b) Energy spectrum as a function of $\protect\mu /t$,
where we have set $J=0.5t$. Red curves indicate the topological edge states,
while cyan curves indicate the bulk states. We have set $L=20$. Color
palette for (a) indicates $\left\vert \protect\psi _{1}\right\vert
^{2}+\left\vert \protect\psi _{L}\right\vert ^{2}$, which is given in Fig.%
\protect\ref{FigDWEne}(d).}
\label{FigMu}
\end{figure}

We show the energy spectrum as a function of $J$ by setting $\mu =0$ in Fig.%
\ref{FigMu}(a). The system is gapped for $J\neq 0$, where topological edge
states emerge at the zero energy. There is a tiny gap around $J=0$\ in Fig.%
\ref{FigMu}(a), which is due to the finite-length effect owing to the
overlap between two topological edge states. We show the energy spectrum as
a function of $\mu $ in Fig.\ref{FigMu}(b), where the\ topological edge
states are found for $\left\vert \mu /t\right\vert <1$.

\subsection{Winding number}

We make a unitary transformation of $H(k)$ by a matrix $U$ given by%
\begin{equation}
U\equiv \left( 
\begin{array}{cccc}
1 & -i & 0 & 0 \\ 
1 & i & 0 & 0 \\ 
0 & 0 & 1 & i \\ 
0 & 0 & 1 & -i%
\end{array}%
\right) .  \notag
\end{equation}%
It follows that%
\begin{equation}
UHU^{-1}=\tau _{0}\otimes H_{0},  \label{TwoCopy}
\end{equation}%
with%
\begin{equation}
H_{0}\equiv \left( 
\begin{array}{cc}
0 & q\left( k\right) \\ 
q^{\ast }\left( k\right) & 0%
\end{array}%
\right) ,
\end{equation}%
and%
\begin{equation}
q\left( k\right) \equiv t\cos k-\mu -iJ\sin k,
\end{equation}%
where we have used the relations $U\Gamma _{0z}U^{-1}=\Gamma _{0x}$ and $%
U\Gamma _{zx}U^{-1}=\Gamma _{0y}$. The Hamiltonian is reduced to a set of
two copies of the two-band\ model $H_{0}$.

There is a hidden $SU(2)$\ symmetry with respect to the first component
in the Hamiltonian (\ref{TwoCopy}),%
\begin{equation}
U_{\text{SU}\left( 2\right) }^{-1}\tau _{0}U_{\text{SU}\left( 2\right)
}=\tau _{0},
\end{equation}%
which protects the condition that the Hamiltonian (\ref{TotalHamil}) is
decomposed into two copies of the Hamiltonian $H_{0}$\ as in Eq.(\ref%
{TwoCopy}).

The topological number is the winding number defined by%
\begin{equation}
w\equiv \frac{i}{2\pi }\int_{0}^{2\pi }q^{-1}\left( k\right) \frac{dq\left(
k\right) }{dk}dk.
\end{equation}%
The winding number is $w=1$ if $q\left( k\right) $ encircles the origin
anticlockwisely in the complex plane made of Re$\left[ q\left( k\right) %
\right] $ and Im$\left[ q\left( k\right) \right] $, while it is $w=-1$ if $%
q\left( k\right) $ encircles the origin clockwisely. On the other hand, it
is $w=0$ if $q\left( k\right) $ does not encircle the origin. We show the
topological phase diagram in the $J$-$\mu $ plane in Fig.\ref{FigPhase}(a).
If $\left\vert \mu /t\right\vert <1$, the winding number is $w=1$ for $J>0$
and $w=-1$ for $J<0$. On the other hand, $w=0$ for $\left\vert \mu
/t\right\vert >1$. It agrees with the number of topological edge states at
one edge in the system $H_{0}$.

\subsection{Comparison to the SSH model}

The Hamiltonian $H_{0}$ is mapped to the long-range SSH\ model $H_{\text{SSH}%
}$\ with $q_{\text{SSH}}\left( k\right) =t_{A}+t_{B}e^{-2ik}-\mu e^{-ik}$ by
making a unitary transformation,%
\begin{equation}
H_{\text{SSH}}=U_{\text{SSH}}H_{0}U_{\text{SSH}}^{-1}\equiv \left( 
\begin{array}{cc}
0 & q_{\text{SSH}}\left( k\right) \\ 
q_{\text{SSH}}^{\ast }\left( k\right) & 0%
\end{array}%
\right) ,
\end{equation}%
where%
\begin{equation}
U_{\text{SSH}}\equiv \left( 
\begin{array}{cc}
1 & 0 \\ 
0 & e^{ik}%
\end{array}%
\right) ,\qquad q_{\text{SSH}}\left( k\right) =q\left( k\right) e^{-ik},
\end{equation}%
and $t_{A}=\left( t-J\right) /2$, $t_{B}=\left( t+J\right) /2$. However,
their topological properties are different, as shown in Figs.\ref{FigPhase}%
(a) and (b). The difference is understood because the momentum-dependent
unitary transformation $U_{\text{SSH}}$ is nonlocal.

Indeed, the winding number is transformed as%
\begin{align}
w_{\text{SSH}}& \equiv \frac{i}{2\pi }\int q_{\text{SSH}}^{-1}\left(
k\right) \frac{dq_{\text{SSH}}\left( k\right) }{dk}dk  \notag \\
& =\frac{i}{2\pi }\int \left( q^{-1}\left( k\right) \frac{dq\left( k\right) 
}{dk}-ik\right) dk=w+1.
\end{align}%
It well explains the relation between these two models as in Fig.\ref%
{FigPhase}(b).

\section{Jackiw-Rebbi zero-energy interface state}

We consider a domain wall configuration $J\left( x\right) $ in the $p$-wave
magnet with length $L$. We assume that it has the same form as in the
conventional magnet, 
\begin{equation}
J\left( x\right) =J\sigma _{z}\tanh \frac{x-L/2-1/2}{\xi },  \label{tanh}
\end{equation}%
satisfying $\lim_{x\rightarrow \infty }J\left( x\right) =J$ and $%
\lim_{x\rightarrow -\infty }J\left( x\right) =-J$, where the magnetization
is flipped without in-plane components. The in-plane component is neglected,
which is a good approximation for a very sharp domain wall, $\xi /L\ll 1$.
We will discuss the effect of the in-plane canting later in the cases of the
Bloch and N\'{e}el domain walls.

The gap closes at $k=\pm \pi /2$, $\mu =0$ and $J=0$. In its vicinity, the
Hamiltonian $H_{0}$ is expanded as%
\begin{equation}
H_{0}=\left( 
\begin{array}{cc}
0 & \mp tk^{\prime }-\mu -iJ \\ 
\mp tk^{\prime }-\mu +iJ & 0%
\end{array}%
\right) ,
\end{equation}%
where we have defined $k^{\prime }=k\mp \pi /2$. The Jackiw-Rebbi equations
read\cite{JR}%
\begin{align}
\{\mp it\partial _{x}-\mu -iJ\left( x\right) \}\psi _{B}\left( x\right) & =0,
\\
\{\mp it\partial _{x}-\mu +iJ\left( x\right) \}\psi _{A}\left( x\right) & =0,
\end{align}%
where $(\psi _{A},\psi _{B})$ is the basis vector of the Hamiltonian $H_{0}$
in the coordinate space.

Case (i): The solution is%
\begin{align}
\psi _{A}\left( x\right) & =0, \\
\psi _{B}\left( x\right) & =\exp \left[ \mp \frac{1}{t}\int^{x}\left(
-J\left( x^{\prime }\right) +i\mu \right) dx^{\prime }\right] ,
\end{align}%
for $J>0$ with $k^{\prime }=k-\pi /2$, and for $J<0$ with $k^{\prime }=k+\pi
/2$.

\begin{figure}[t]
\centerline{\includegraphics[width=0.48\textwidth]{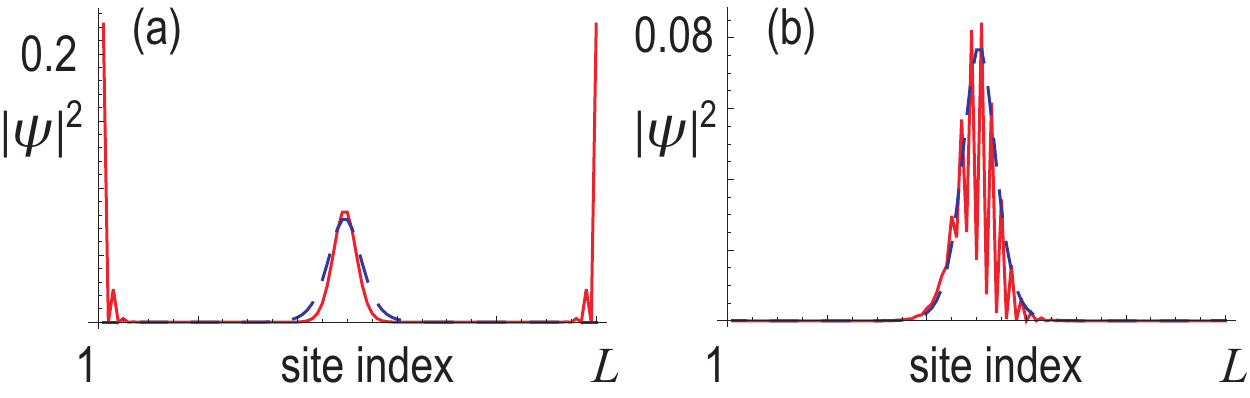}}
\caption{(a) Spatial profile of the topological interface and edge states
base on the Jackiw-Rebbi model. (b) Spatial profile of the topological
interface states based on the magnetic domain-wall model. Red curves
indicate numerical results based on the tight-binding model, while blue
dashed curves indicate analytic solutions based on the Jackiw-Rebbi
model. We have set $\protect\xi =5$ and $J=0.5t$. We have use a chain wit
the length $L=100$.}
\label{FigJR}
\end{figure}

Case (ii): The solution is%
\begin{align}
\psi _{A}\left( x\right) & =\exp \left[ \mp \frac{1}{t}\int^{x}\left(
J\left( x^{\prime }\right) +i\mu \right) dx^{\prime }\right] , \\
\psi _{B}\left( x\right) & =0,
\end{align}%
for $J<0$ with $k^{\prime }=k-\pi /2$, and for $J>0$ with $k^{\prime }=k+\pi
/2$.

In the case (i), with the use of Eq.(\ref{tanh}), we obtain 
\begin{equation}
\psi _{B}\left( x\right) =\sqrt{\frac{\Gamma \left( 1/2+\left\vert
J\right\vert \xi /t\right) }{\sqrt{\pi }\xi \Gamma \left( \left\vert
J\right\vert \xi /t\right) }}e^{i\mu x/t}\cosh ^{-\frac{\left\vert
J\right\vert \xi }{t}}\frac{x-\frac{L+1}{2}}{\xi }.  \label{WaveB}
\end{equation}%
In the case (ii), we obtain Eq.(\ref{WaveB}) for $\psi _{A}\left( x\right) $%
. We show the spatial profile $\left\vert \psi \left( x\right) \right\vert
^{2}\equiv \left\vert \psi _{A}\left( x\right) \right\vert ^{2}+\left\vert
\psi _{B}\left( x\right) \right\vert ^{2}$ by a blue curve in Fig.\ref{FigJR}%
(a).

We also show the spatial profile $\left\vert \psi \left( x\right)
\right\vert ^{2}$ obtained numerically based on the tight-binding model by a
red curve in Fig.\ref{FigJR}(a). The tight-binding result consists of the
topological interface state at $x=L/2$ and the topological edge states at $%
x=1$, $L$. There are four-fold degenerate topological interface states from $%
k=\pm \pi /2$\ and from the two copies of the Hamiltonian $H_{0}$.

One interface state has a half charge%
\begin{equation}
\rho \left( x\right) =\frac{e}{2}\left\vert \psi \left( x\right) \right\vert
^{2},
\end{equation}%
according to the standard argument of the Jackiw-Rebbi state\cite{JR}. The
charge is localized at the domain wall.

\subsection{Bloch and N\'{e}el magnetic domain walls}

The domain wall is actually canted and forms the Bloch or the N\'{e}el
domain wall instead of the Jackiw-Rebbi type domain wall. The magnetization
is%
\begin{equation}
J\left( \sigma _{z}\cos \theta \left( x\right) +\sigma _{x}\sin \theta
\left( x\right) \right)  \label{Neel}
\end{equation}%
for the N\'{e}el domain wall, and 
\begin{equation}
J\left( \sigma _{z}\cos \theta \left( x\right) +\sigma _{y}\sin \theta
\left( x\right) \right)  \label{Bloch}
\end{equation}%
for the Bloch domain wall, where we have set%
\begin{equation}
\cos \theta \left( x\right) =\tanh \frac{x-1/2}{\xi },\quad \sin \theta
\left( x\right) =\text{sech}\frac{x-1/2}{\xi }.
\end{equation}

\begin{figure}[t]
\centerline{\includegraphics[width=0.44\textwidth]{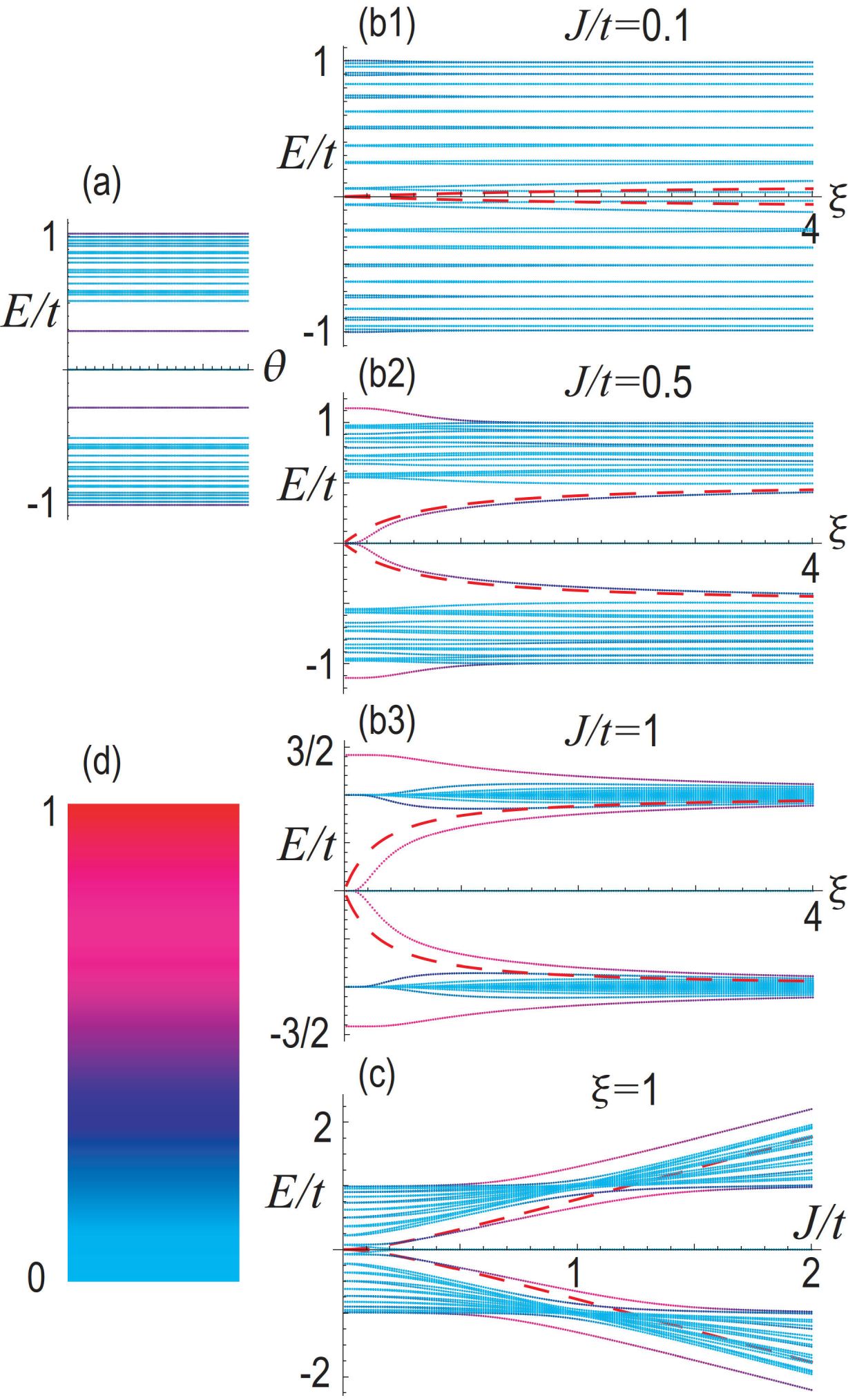}}
\caption{(a) Energy spectrum as a function of $\protect\theta $. We have
set $J=0.5t$, $\protect\xi =1$\ and $\protect\mu =0$. (b1), (b2) and (b3)
Energy spectrum as a function of $\protect\xi $. The red dashed curves
indicate analytic solutions (\protect\ref{E1}) based on the first-order
perturbation theory. We have set (b1) $J=0.1t$, (b2) $J=0.5t$\ and (b3) $J=t$
(c) Energy spectrum as a function of $J/t$ . We have set $\protect\mu =0$.
Red curves indicate the interface states, while cyan curves indicate the
bulk or edge states. The length is $L=20$. (d) Color palette indicates $%
\left\vert \protect\psi _{1}\right\vert ^{2}+\left\vert \protect\psi %
_{L}\right\vert ^{2}$.}
\label{FigDWEne}
\end{figure}

The energy spectrum is identical between the Bloch and N\'{e}el domain
walls as shown in Fig.\ref{FigDWEne}(a). It is understood that these two
systems are connected via unitary transformation,%
\begin{equation}
H_{\text{Bloch}}=U^{-1}H_{\text{N\'{e}el}}U,
\end{equation}%
where%
\begin{equation}
U\equiv \exp \left[ \frac{i\pi }{4}\sigma _{z}\right]
\end{equation}%
is the rotation along th $z$ axis.

We estimate the energy of the Bloch and N\'{e}el domain walls. We use the
first-order perturbation theory\cite{Wakatsuki} in the presence of the term
proportional to $\sin \theta \left( x\right) $ in Eqs.(\ref{Neel}) and (\ref%
{Bloch}), where the eigenfunctions do not change but the energy is shifted.
The first-order energy displacement is calculated with the use of Eq.(\ref%
{WaveB}) as%
\begin{equation}
E=\int \left\vert \psi \left( x\right) \right\vert ^{2}\sin \theta \left(
x\right) dx=J\frac{\Gamma ^{2}\left( \frac{1}{2}+\frac{J\xi }{t}\right) }{%
\Gamma \left( 1+\frac{J\xi }{t}\right) \Gamma \left( \frac{J\xi }{t}\right) }%
,  \label{E1}
\end{equation}%
which is independent of the canted direction between the Bloch and the N\'{e}%
el domain walls.

We show the energy (\ref{E1}) as a function of $\xi $ by dotted red curves
in Fig.\ref{FigDWEne}(b1), (b2) and (b3). They well agree with the numerical
results on the energy spectrum in Fig.\ref{FigDWEne}(b1), (b2) and (b3),
where the first-order perturbation theory and the continuum theory are not
assumed. There is a tiny discrepancy at small $\xi $ between the analytical
and the numerical results. The numerical result shows a flat dispersion for $%
\left\vert J\xi /t\right\vert \ll 1$, while it is linear in the analytic
result. It is due to the break down of the continuum approximation. For
small $\xi $, Eq.(\ref{E1}) yields%
\begin{equation}
E=J\pi \xi /t+o\left( \left( J\pi \xi /t\right) ^{2}\right) .
\end{equation}%
On the other hand, for large $\xi $, Eq.(\ref{E1}) yields%
\begin{equation}
E/J=1-t/4J\xi +\left( t/4J\xi \right) ^{2}/2+o\left( \left( t/4J\xi \right)
^{3}\right) .
\end{equation}

We show the energy spectrum as a function of $J/t$\ in Fig.\ref{FigDWEne}%
(c). It shows that the perturbation theory is valid for $J/t<1$. $J/t$\ is
much smaller than 1 in actual materials, which ensures the perturbation
theory.

The interface states persist for the Bloch and the N\'{e}el domain walls but
acquire nonzero energy as shown in Fig.\ref{FigJR}(b), whose spatial profile
is well coincident with that of the Jackiw-Rebbi solution as depicted by a
blue curve. It is understood as follows. The contribution of the in-plane
component is tiny for a sharp domain wall. In this regime, we can treat the
in-plane Hamiltonian as a perturbation term. Hence, the eigenstates remains
as they are but the energy is modified in the first-order perturbation
theory.

A comment is in order. The domain wall width is controlled by introducing
the easy-axis anisotropy term $-AS_{z}^{2}$ to the system, which favors the
spin pointing up or down. The domain wall width becomes smaller for larger
easy-axis anisotropy.

\section{2D model}

\begin{figure}[t]
\centerline{\includegraphics[width=0.48\textwidth]{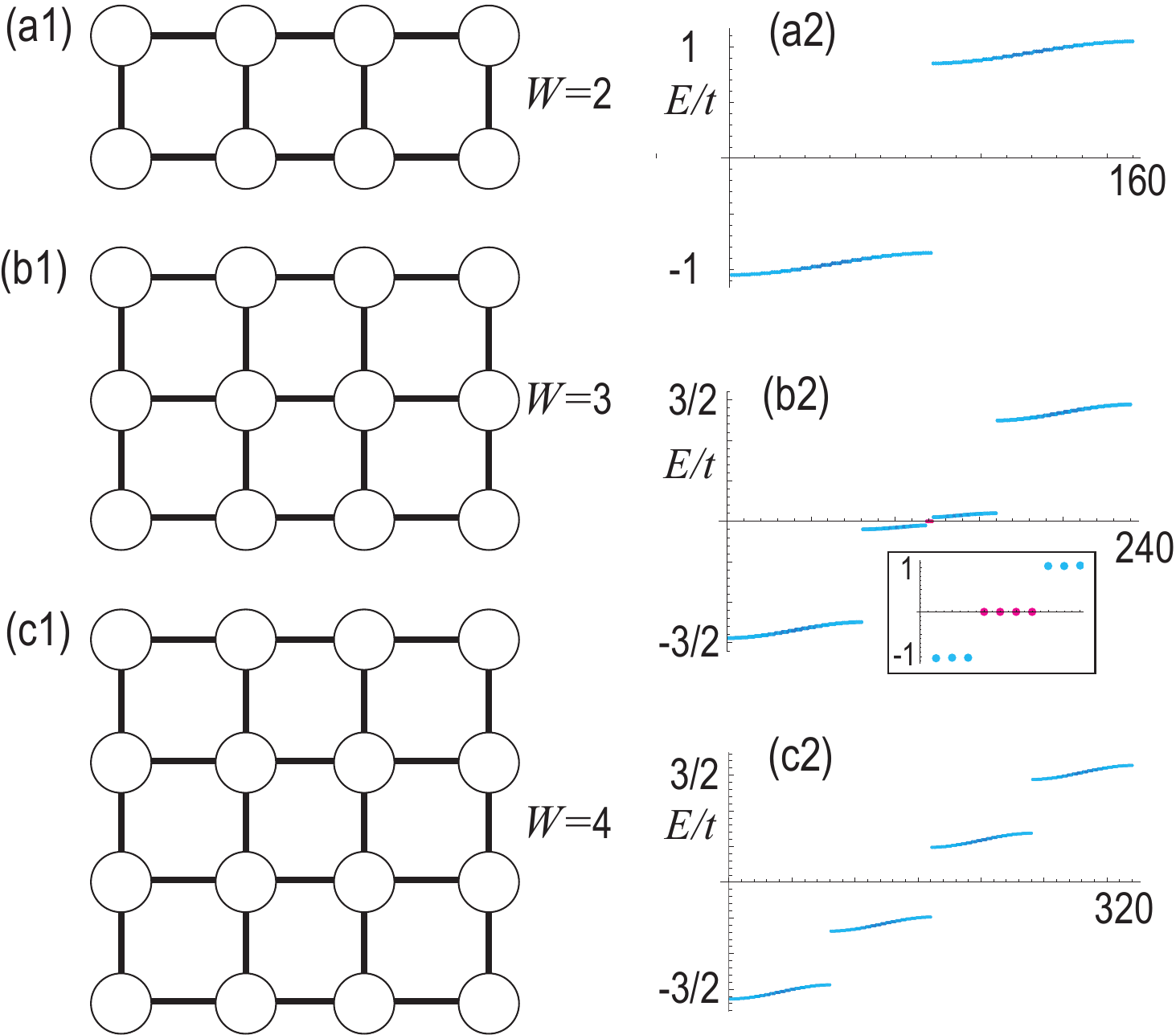}}
\caption{Illustration of (a1) two-leg ladder model, (b1) three-leg ladder
model and (c1) four-leg ladder model. (a2), (b2) and (c2)$.$\ Energy
spectrum for a finite chain with\ the width\ $W$. The horizontal axis
is the site index. (a2) Nanoribbon described by Eq.(\protect\ref{2D}) with
the width $W=2$, (b2) Nanoribbon with the width $W=3$ and (c2)
Nanoribbon with the width $W=4$. Insets in (b2) show the enlarged
energy spectrum in the vicinity of the zero energy, where red dots indicate
the topological edge states, while cyan dots indicate the bulk states.
Topological edge states emerge only for (b2). We have set $J=0.5t$, $%
\protect\mu =0$ and $L=20$. Color palette for (a) indicates $\left\vert 
\protect\psi _{1}\right\vert ^{2}+\left\vert \protect\psi _{L}\right\vert
^{2}$, which is given in Fig.\protect\ref{FigDWEne}(d).}
\label{FigEdgeW}
\end{figure}

An actual sample has a finite width $W$. We consider a two-dimensional
generalization of the model, where the Hamiltonian is given by%
\begin{equation}
H_{2}=\left( t\cos k_{x}+t\cos k_{y}-\mu \right) \Gamma _{0z}+\left( J\sin
k_{x}\right) \Gamma _{zx}.  \label{2D}
\end{equation}%
It is constructed from the one-dimensional model (\ref{TotalHamil}) with the
replacement of the chemical potential $\mu $\ by $\mu \left( k_{y}\right)
=\mu -t\cos k_{y}$. The topological class of the class BDI in the two
dimensions is trivial\cite{Ryu}. Nevertheless, there emerge topological edge
states in the ladder geometry with finite width $W$\ as illustrated in Fig.%
\ref{FigEdgeW}(a1), (b1) and (c1). Note that the ladder system with finite
width is a one-dimensional system and belongs to the class $Z$.

We study a nanoribbon geometry based on the Hamiltonian (\ref{2D}). We find
that there emerge 4 zero-energy topological edge states for odd width
nanoribbons. Examples for the width $W=2,3,4$ are shown in Fig.\ref{FigEdgeW}%
(a2), (b2) and (c2).

This even-odd effect with respect to $W$ is understood as follows. The
Hamiltonian is actually made of two copies as in Eq.(\ref{TwoCopy}). In the
vicinity of the zero energy, each Hamiltonian is well approximated by two
chains with length $W$\ at $x=1$ and $L$, each of which\ is composed of
topological edge states. The effective Hamiltonian of one chain is given by%
\begin{equation}
H_{\text{eff}}=t\sum_{y=1}^{W-1}\left( \hat{f}_{y}^{\dagger }\hat{f}_{y+1}+%
\hat{f}_{y+1}^{\dagger }\hat{f}_{y}\right) ,
\end{equation}%
for $W\geq 2$, where $\hat{f}_{y}$ is an annihilation operator of the
topological edge state. The energy spectrum is obtained as $E_{\text{eff}%
}\left( j\right) =t\cos \left( 2\pi j/W+1\right) $ with $j=1,\cdots W$.
There is one zero-energy state for odd $W$\ and no zero-energy state for
even $W$. As a result, by considering two copies of the model and two edges,
we have 4 zero-energy states for a nanoribbon with odd $W$.

\begin{figure}[t]
\centerline{\includegraphics[width=0.44\textwidth]{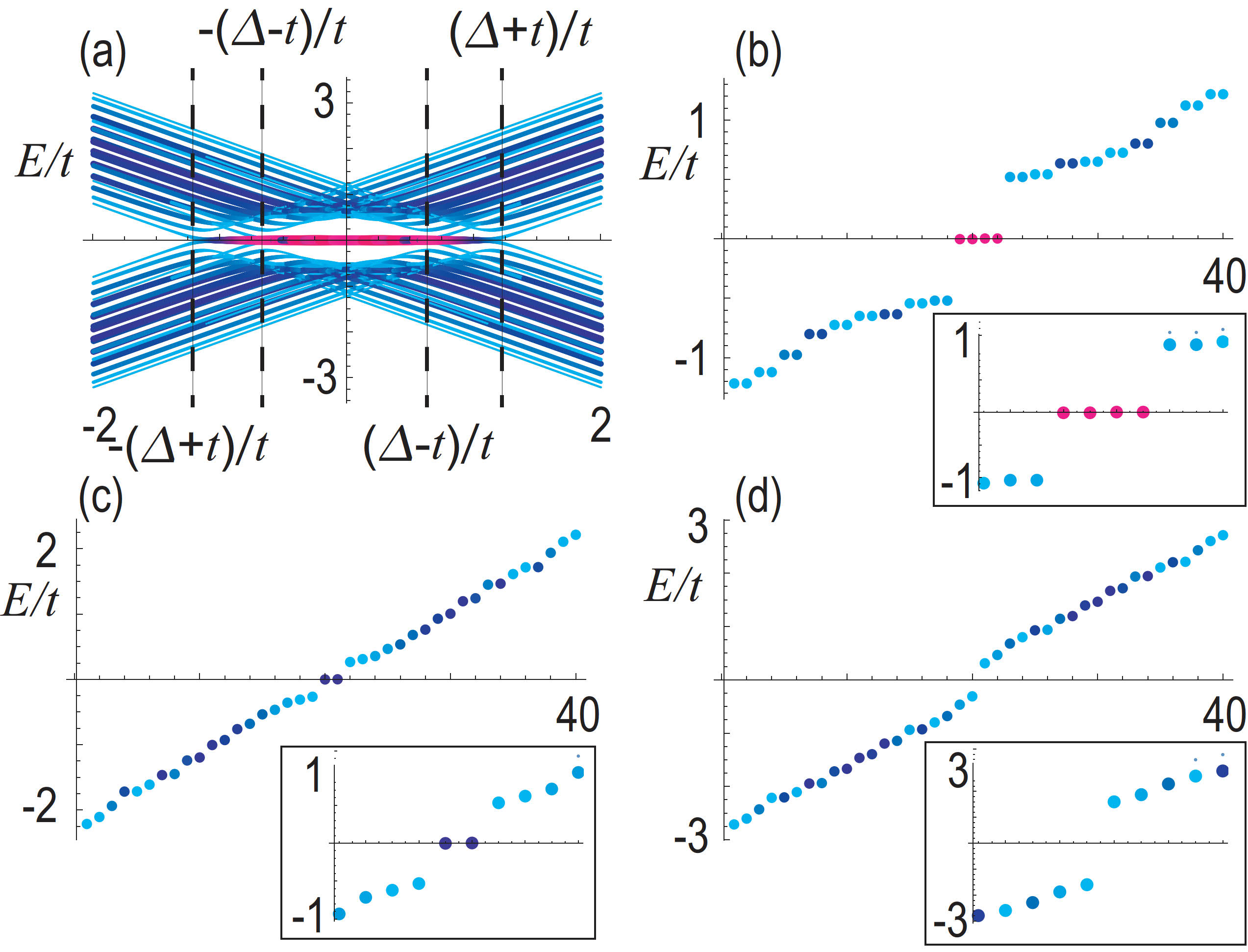}}
\caption{(a) Energy spectrum as a function of $\protect\mu /t$.\ We have
set $J=0.5t$, $\protect\xi =1$ and $\Delta =0.5t$. (b) Energy spectrum as a
function of the state index with $\protect\mu =0$, where there are four
zero-energy edge states. (c) $\protect\mu =t$, where there are two
zero-energy edge states. (d) $\protect\mu =2t$, where there are no
zero-energy edge states. We have used a chain with the length $L=10$. Color
palette for (a) indicates $\left\vert \protect\psi _{1}\right\vert
^{2}+\left\vert \protect\psi _{L}\right\vert ^{2}$, which is given in Fig.%
\protect\ref{FigDWEne}(d).}
\label{FigMajo}
\end{figure}

A comment is in order. There are even-odd effects of the emergence of the
edge or interface depending on the width of the ladder. It requires atomic
scale layer control to observe the phenomenon in experiments.

\section{Majorana fermions}

Majorana fermions are studied in the case of $d$-wave altermagnets\cite%
{Zu2023,Li2023,Gho}. We study Majorana fermions in $p$-wave magnets. We
consider the case where $s$-wave superconducting paring between up and down
spins with opposite orbitals is introduced via proximity coupling,%
\begin{equation}
H_{\Delta }=\Delta \sum_{x,\tau =\pm 1}\left( c_{x,\uparrow ,\tau
}c_{x,\downarrow ,-\tau }+c_{x,\downarrow ,-\tau }^{\dagger }c_{x,\uparrow
,\tau }^{\dagger }\right).
\end{equation}%
\ The Bogoliubov-de Gennes Hamiltonian is given by%
\begin{eqnarray}
H_{\text{BdG}} &=&\left( 
\begin{array}{cc}
H_{\uparrow }\left( k\right) & \Delta \tau _{x} \\ 
\Delta \tau _{x} & -H_{\downarrow }\left( -k\right)%
\end{array}%
\right)  \notag \\
&=&\left[ \left( t\cos k_{x}-\mu \right) \tau _{z}+J\tau _{x}\sin k_{x}%
\right] \otimes \zeta _{z}  \notag \\
&&+\Delta \tau _{x}\otimes \zeta _{x}.  \label{HBdG}
\end{eqnarray}%

The energy spectrum is%
\begin{equation}
E_{\text{BdG}}=\pm \sqrt{J^{2}\sin ^{2}k_{x}+\left( t\cos k_{x}-\mu \pm
\Delta \right) }.
\end{equation}%
The gap closes at $k_{x}=0$ under the condition%
\begin{equation}
t-\mu \pm \Delta =0.
\end{equation}%
We show the energy spectrum as a function of $\mu $\ in Fig.\ref{FigMajo}%
(a). There are four Majorana zero-energy states for $\left\vert \mu
\right\vert <t-\Delta $\ as shown in Fig.\ref{FigMajo}(c) and there are two
Majorana zero-energy states for $t-\Delta <\left\vert \mu \right\vert
<t+\Delta $\ as shown in Fig.\ref{FigMajo}(c). There is no zero-energy
states for $\left\vert \mu \right\vert >t+\Delta $\ as shown in Fig.\ref%
{FigMajo}(d).

\begin{figure}[t]
\centerline{\includegraphics[width=0.44\textwidth]{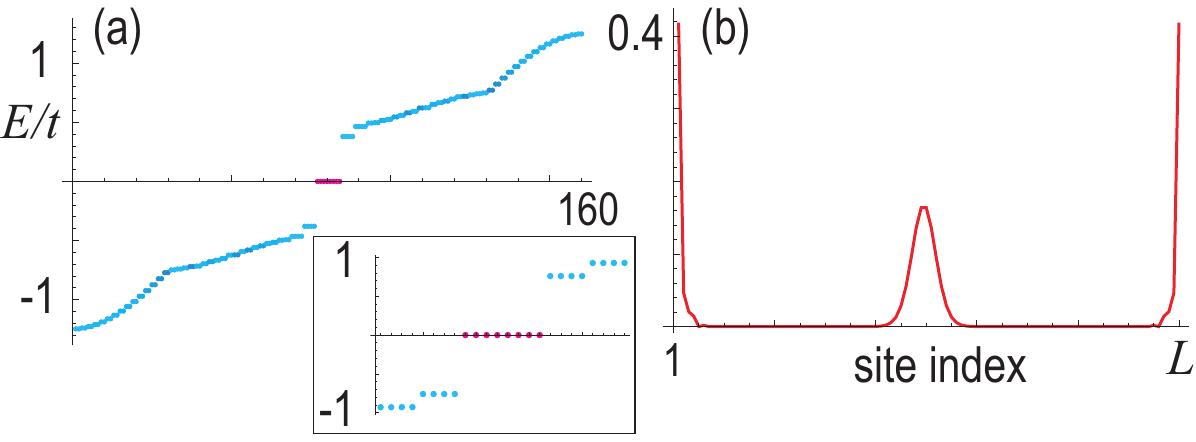}}
\caption{(a) Energy spectrum with a $p$-wave magnetic domain wall. The
inset shows an enlarged figure in the vicinity of the zero energy, where
there are eight zero-energy states. They are four edge states and four
interface states. (b) Spatial profile of Majorana Jackiw-Rebbi mode. We have
used a chain with the length $L=40$. We have set $J=0.5t$, $\protect\mu =0$, 
$\protect\xi =5$ and $\Delta =0.5t$. Color palette for (a) indicates $%
\left\vert \protect\psi _{1}\right\vert ^{2}+\left\vert \protect\psi %
_{L}\right\vert ^{2}$, which is given in Fig.\protect\ref{FigDWEne}(d).}
\label{FigMajoJR}
\end{figure}

It is understood as follows. Actually, the Hamiltonian (\ref{HBdG}) is
unitary equivalent to two copies of the Kitaev topological superconductor
model, 
\begin{equation}
U_{\text{BdG}}H_{\text{BdG}}U_{\text{BdG}}^{-1}=\left( 
\begin{array}{cc}
H_{1} & 0 \\ 
0 & H_{2}%
\end{array}%
\right)
\end{equation}%
with%
\begin{eqnarray}
H_{1} &=&\left( t\cos k_{x}-\mu +\Delta \right) \sigma _{z}+J\sin
k_{x}\sigma _{x}, \\
H_{2} &=&\left( t\cos k_{x}-\mu -\Delta \right) \sigma _{z}+J\sin
k_{x}\sigma _{x},
\end{eqnarray}%
where the unitary transformation is given by%
\begin{equation}
U_{\text{BdG}}=\left( 
\begin{array}{cccc}
1 & 0 & 0 & 1 \\ 
0 & -1 & 1 & 0 \\ 
1 & 0 & 0 & -1 \\ 
0 & 1 & 1 & 0%
\end{array}%
\right) ,
\end{equation}%
where we have used the relations $U\Gamma _{zz}U^{-1}=\Gamma _{0z}$, $%
U\Gamma _{xz}U^{-1}=\Gamma _{0x}$ and $U\Gamma _{xx}U^{-1}=\Gamma _{zz}$.

The unit vector defined by%
\begin{equation}
\mathbf{n}_{j}\left( k\right) \equiv H_{j}\left( k\right) /\left\vert
H_{j}\left( k\right) \right\vert
\end{equation}%
points the $z$ direction at $k=0$ and $\pi $ due to particle-hole symmetry.
Then, we can define 
\begin{equation}
n_{j}\left( k\right) \sigma _{z}=H_{j}\left( k\right) /\left\vert
H_{j}\left( k\right) \right\vert
\end{equation}%
for $k=0$ and $\pi $. As in the case of the Kitaev model, the topological
index is the Z$_{2}$ index defined by%
\begin{equation}
\nu \equiv n_{j}\left( 0\right) n_{j}\left( \pi \right) .
\end{equation}%
The system is trivial for $\nu =1$, while the system is topological for $\nu
=-1$. The Hamiltonian $H_{1}$\ is topological for $\left\vert \mu
\right\vert <t-\Delta $, while it is trivial for $\left\vert \mu \right\vert
>t-\Delta $. On the other hand, the Hamiltonian $H_{2}$\ is topological for $%
\left\vert \mu \right\vert <t+\Delta $, while it is trivial for $\left\vert
\mu \right\vert >t+\Delta $. It well explains the result shown in Fig.\ref%
{FigMajo}(d) because there emerge two Majorana edge states for each model in
the topological phase. 

The energy spectrum with the $p$-wave magnetic domain wall is shown in
Fig.\ref{FigMajoJR}(a). The spatial profile of the zero-energy states is
shown in Fig.\ref{FigMajoJR}(b), where the Majorana interface states are
formed at the domain wall. 

\section{Discussions}

We have investigated physical properties of $p$-wave magnets. The $p$%
-wave magnets are antiferromagnets just as altermagnets are. The N\'{e}el
vector is observable by way of anomalous Hall effects in altermagnets. This
is not the case in $p$-wave magnets. To overcome this, we have explored the
topological property. We have argued that the domain wall position will be
detectable and controllable by electric field because it is a Jackiw-Rebbi
state. Just as altermagnets have zero-net magnetization, $p$-wave magnets
have zero-net magnetization.

Electron-electron interactions are neglected in the present study. There
are several studies on interaction effects in the SSH model\cite%
{WChen,Sirk,Marq,Yah,Jin,Melo,Salvo}. The topological edge states are robust
under the critical strength of electron-electron interactions, which is much
smaller than that in ordinary materials. Then, it is adequate to neglect
electron-electron interactions to study topological phases. 

This work is supported by CREST, JST (Grants No. JPMJCR20T2) and
Grants-in-Aid for Scientific Research from MEXT KAKENHI (Grant No.
23H00171). 

\end{document}